\documentclass[journal=mamobx,manuscript=article]{achemso}
\usepackage[version=3]{mhchem} 
\usepackage{graphicx}
\usepackage{caption}
\usepackage{subcaption}

\author{Bekele Gurmessa}
\author{Andrew B. Croll}
\affiliation[NDSU]{Department of Physics, North Dakota State University}
\email {andrew.croll@ndsu.edu}

\title {The Influence of Thin Film Confinement on Surface Plasticity in Polystyrene and Poly(2-vinylpyridine) Homopolymer and Block Copolymer Films }


\begin{document}


\begin{abstract}
Thin block copolymer films have attracted considerable academic attention because of their ability to self-assemble into various microstructures, many of which have potential technological applications.  Despite the ongoing interest, little effort has focused on the onset of plasticity and failure which are important factors for the eventual adoption of these materials.  Here we use delamination to impart a quantifiable local stain on thin films of homopolymer polystyrene and poly(2-vinylpyridine), as well as block copolymers made of styrene and 2-vinylpyridine.  Direct observation of the damage caused by bending with atomic force microscopy and laser scanning confocal microscopy, leads to the identification of a critical strain for the onset of plasticity.  Moving beyond our initial scaling analysis, the more quantitative analysis presented here shows strain levels for thick films to be comparable to bulk measurements.  Monitoring the critical strain leads to several observations: 1.) as-cast PS-P2VP has low critical strain, 2.) annealing slowly increases critical strain as microstructural ordering takes place, and 3.) similar to the homopolymer, both as cast and ordered films both show increasing critical strain under confinement.  
\end{abstract}

\section{Introduction}

Block copolymers are facinating materials with a richness of mechanical behavior emerging from the organized nanostructures which can form within.~\cite{ Bates1990, Lewis1969, Schwier1985, Honeker1996, Weidisch1999a, Weidisch1999b, Weidisch2000, Cohen2000, Lach2004, Ye2013, Ryu2002, Ruokolainen2002, Lee2005, Kim2006, Makke2012}.  Over the past several decades, many block copolymer systems have been industrialized and much has been learned about the details of their bulk mechanical behavior.  Studies have largely focused on hard-soft material combinations, driven by attempts to toughen materials while maintaining other desirable properties (optical transparency, for example)~\cite{Schwier1985,Honeker1996, Weidisch1999a, Weidisch1999b, Weidisch2000, Cohen2000, Lach2004, Ye2013}.  There is also some precedent for the study of hard-hard systems in attempts to compatiblize homopolymer mixtures or attempts to add crystallinity.~\cite{Ryu2002, Ruokolainen2002, Lee2005, Kim2006}  As one might imagine, material properties are found to depend on almost all of the features governing the nanostructure of the materials (volume fraction\cite{Weidisch1999b,Weidisch2000,Lach2004}, chain-chain interactions\cite{Honeker1996, Weidisch2001}, chain orientation and alignment\cite{Weidisch2000, Cohen2000, Ruokolainen2002, Ye2013}, thermal history\cite{Ruokolainen2002}).

Currently, the renewed industrial drive towards miniturization has lead to increased attention on block copolymer in thin film geometries.~\cite{Matsen1997,Fasolka2001,Segalman2005}  Despite the interest, there remains little direct mechanical characterization of block copolymer thin films, largely due to the lack of established methods for the mechanical testing of thin polymer films in general.  The result is a gap in knowledge of properties that should be expected to vary from their bulk values, or at the least be much more severely constrained by geometric and interfacial phenomena.  In an attempt to fill this gap, we have used a recently-developed localized-bending methodology to examine the early stages of plastic failure in a model block copolymer, polystyrene-b-poly(2-vinylpyridine) or PS-P2VP, and the two homopolymers used to make it up (polystyrene, PS, and poly(2-vinylpyridine), P2VP).\cite{Gurmessa2013}  An increase in the critical strain for plastic deformation was observed in all samples as film thickness is reduced.  The critical strain recorded for the symmetric block copolymer fell halfway between the values recorded for the two homopolymers, but only after sufficient annealing for significant microphase separation to take place.

The glass transition temperature, $T_{g}$, is a prime example of a material property that has often been observed to significantly deviate from its bulk values as a polymer thin film's thickness drops below $\sim 100$~nm.~\cite{Keddie1994, Sharp2003, Baumchen2012, Ellison2003, Cheng2007, Roth2007a, Roth2007b, Fakhraai2008, Keewook2011, Pye2011, Lang2014, Torres2000, Varnik2002, Zhang2011}  The changes are known to be related to the environment experienced by the film's surfaces.  Surface affinity leads to slower than usual dynamics and an increased $T_{g}$, while low affinity or free surfaces generally lead to increased dynamics.\cite{Keddie1994,Sharp2003,Baumchen2012}  For example, PS on a silicon support is found to have a glass transition that decreases upon confinement, whereas P2VP is found to have a significantly increased glass transition temperature due to increased hydrogen bonding with the substrate.\cite{Keddie1994, Roth2007a}  On the other hand, free surfaces generally lead to universal decreases in $T_{g}$.\cite{Sharp2003,Baumchen2012}  It is important to note that recent work with stacked films \cite{Roth2007a, Roth2007b}, and soft substrates \cite{Zhang2011,Lang2014} has shown that the surface chemistry alone is not enough to completely explain measurements.

Given the relationship between internal dynamics and mechanical behavior in bulk materials, it is not surprising that many researchers have attempted to find thin film effects in the mechanical properties of polymer films\cite{Lee1996, Yoshimoto2005, Torres2009, Lee2012, Chung2014,Cheng2007, Fakhraai2008, OConnell2005, Tweedie2007, Stafford2006, Zhao2000, Bohme2002, Miyake2006, Forrest1998, Crosby2015}.  The results have been mixed.  For example, stiffining has been observed\cite{OConnell2005}, moduli have been observed to increase\cite{Tweedie2007}, decrease\cite{Zhao2000, Bohme2002, Miyake2006, Stafford2006} or remain unchanged in experiments\cite{Forrest1998, Crosby2015}.  While there is no clear explanation for the discrepancy, different substrates and annealing histories may play a role, as well as differences in the lengthscales and timescales probed in the different measurements.  Large deformations may be influenced by changes in the inter-chain entanglement density that arise due to confinement \cite{Brown1996, Si2005}, whereas small deformations are not.\cite{Gurmessa2013}  Slow, low-amplitude and direct mechanical experiments may be the only way to link with $T_{g}$ phenomena, but even this is unclear.\cite{Torres2009,Schnell2011,Lang2014}

\begin{figure}
\centering
\includegraphics[height=0.5\textwidth]{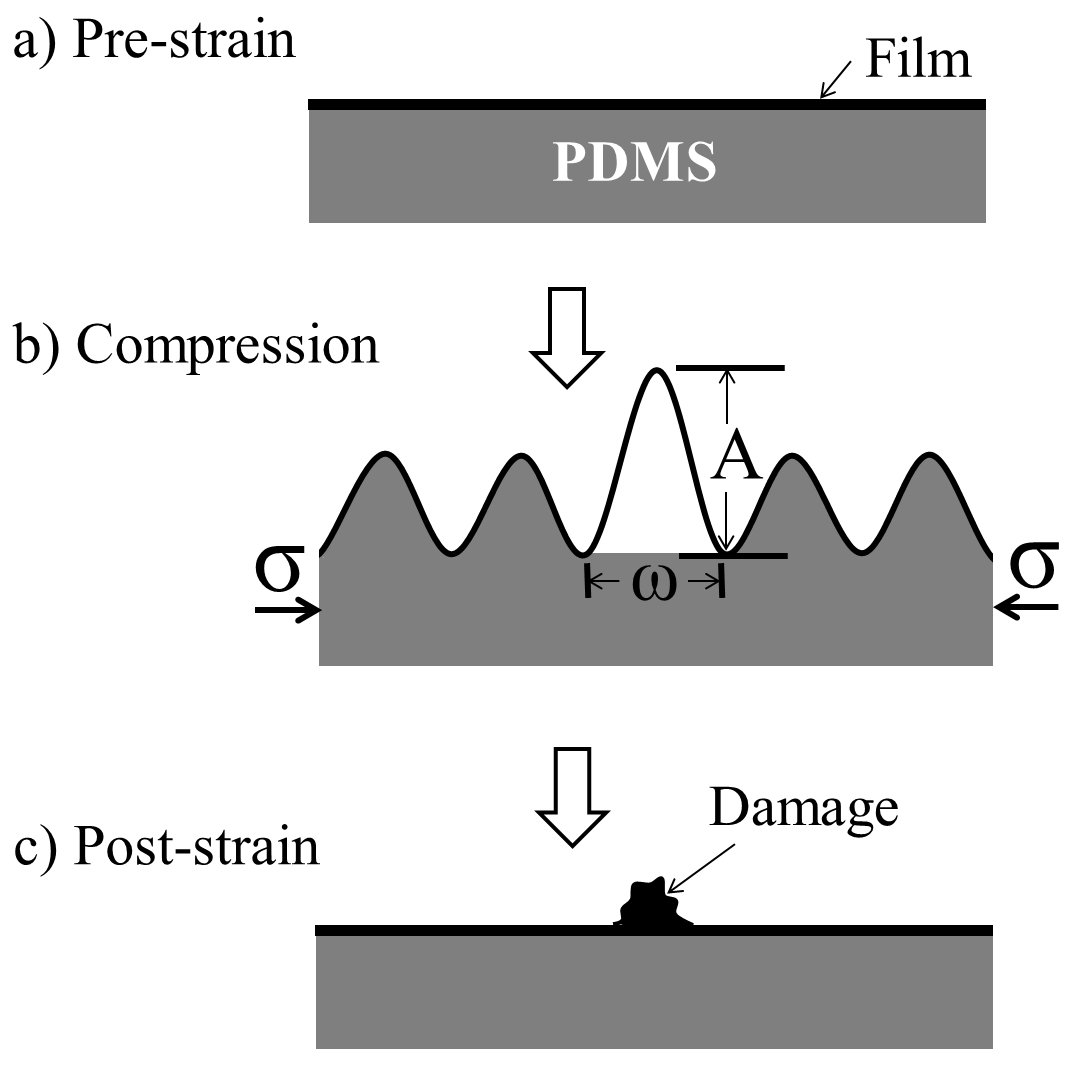}
\caption{schematic of the experimental geometry.  (a) A thin polymer film is laminated to a soft substrate.  (b) The composite is compressed causing the film to wrinkle and delaminate.  (c) Compression is removed and the film returns to a flat state, although damage may have occurred at locations of high curvature. }\label{scheme1}
\end{figure}

Recently, we have used delamination as a method to create simple, localized bending in thin free-standing polystyrene films.\cite{Gurmessa2013}  In these experiments a thin polymer film is mounted to a soft substrate and the composite is slowly compressed.  Elastic instability leads to repetitive buckling (wrinkling) or delamination of the film from the substrate.  The bending in a delaminated section of film is easily quantified by Laser Scanning Confocal Microscopy (LSCM) or Atomic Force Microscopy (AFM) and local curvature (thus strain) can be directly evaluated.  More importantly, the film can be returned to its initially flat state by the removal of compression in the composite.  The flat film can then be interrogated with AFM or LSCM in order to identify signs of yield.  Damage can then be correlated with the surface strain created by the bending in the preceeding delmaination (see Fig~\ref{scheme1}).  The result is important because it evaluates a well defined physical endpoint of a mechanical test (failure) and relies only on tracking the film's position in space.  Hence, regardless of the details of what a modulus might \textit{mean} for a heterogeneous thin-film, a clear outcome of the deformation can be mapped.

The goal of this work is threefold.  First, we wish to examine the onset of plasticity in additional materials in order to prove that we are accurately measuring a material-dependent property.  Secondly, we wish to improve our initial scaling analysis to the point where we can tentatively claim quantitative results.  Finally, and most importantly, we wish to explore how microstructure might affect the failure process in thin block copolymer films.  The paper is organized largely along these lines.  After experimental procedures are described in detail, results for PS and P2VP hompolymer are discussed.  Lastly, results observed with PS-P2VP block copolymer are described and brief conclusions are made.

\section{Experimental}

\textbf{Materials}: Monodisperse, symmetric and asymmetric  polystyrene-b-poly(2-vinylpyridine) (PS-P2VP) were purchased from Polymer Source Inc. and used as received.  The symmetric PS-P2VP has a PS block of molecular weight $M_n$ = 40 kg/mol and a P2VP block of $M_n$ = 40.5 kg/mol. The polydispersity index (PDI) of the diblock is 1.02. The second molecule considered was a cylinder forming PS-P2VP with PS block of $M_n$ = 56~kg/mol and a P2VP block of $M_n$=  21 kg/mol, and has PDI of 1.06.  The two molecules were chosen to have differing microphases but similar total molecular weight. In addition, polystyrene homopolymer of molecular weights $M_n = 1.3$~Mg/mol and $M_n = 1.2$~kg/mol with PDI of 1.15 and 1.04 respectively and  Poly(2-vinylpyridine) (P2VP) homopolymer of molecular weight $M_n = 135$~kg/mol and PDI of 1.06, also  purchased from Polymer Source Inc., were used in the experiment.  

\textbf{Substrate and Film Preparation}: The substrate was prepared by mixing  PDMS prepolymer and cross-linker (Dow chemical Corning, Sylgard 184) in a 20:1 weight ratio, respectively. The mixture was then degassed and poured into a petri dish, cured at $85$~$^\circ$C for two hours and then left in vacuum oven for the next 12-15 hours. Finally, after the PDMS is cooled down, it is cut into a rectangular sections of dimension $3$~mm~$\times~12$~mm~$\times~70$~mm.

Matching pairs of samples were made, by spin casting PS-b-P2VP/toluene solution onto a freshly cleaved mica substrate. The concentration of the polymer (by weight) considered here ranges from $0.5$\% - $3$\%.  The solvent was toluene (Anhydrous 99.8\%, Sigma-Aldrich Inc.) and the solutions were filtered (pore size 0.482 $\mu$m, Cadence Science Inc.) before spin coating.  The first sample is used as-cast and the second sample undergoes a step of annealing to enable self-assembly of the block copolymer microdomains. To initiate the self-assembly process,  samples were annealed at temperatures ranging from $165$ $^\circ$C - $195$ $^\circ$C - all well above the glass transition temperature of the bulk PS-b-P2VP ($\sim 100$ $^\circ$C) either in an air environment or in a glove box filled with dry nitrogen.  Finally, samples are transferred to a clean deionized water surface (Milli-Q) and subsequently transferred to a PDMS substrate loaded on a home built strain stage in order to impart a compressive stress. In all cases the films are imaged with Laser Scanning Confocal Microscopy (LSCM - Olympus Fluoview 1000) or with Atomic Force Microscopy (AFM - DI Dimension 2100).

\textbf{Thickness Measurement}: Film thickness is one of the dominant geometric properties related to bending and must be measured carefully.  The homopolymer and as-cast block copolymer present little challenge in analysis; films are scratched near the feature of interest and Atomic Force Microscopy (AFM - DI Dimension 2100) is used to locally measure a film thickness.  However, when the thin film is ordered lamella-forming block copolymer will be decorated by terraces (islands or holes) whenever the as-cast thickness is not commensurate with the lamellar thickness. Under such conditions, using AFM cross-section analysis alone is insufficient to completely describe a thickness; an average thickness must be contrived from the density of surface features, $f$, the lamellar thickness $L_0$, the number of complete layers below the surface, $n$, and volume conservation.  Specifically, we calculate the average thickness as,
\begin{equation}\label{avgthickness}
h_{avg} = f \times L_0 + n \times L_0 .
\end{equation}
Alternatively, the largest thickness $(n+1 )L_0$ and thickness of only complete layers, $nL_0$ was considered in analysis.  Neither resulted in significant changes in the overall trends observed.  With the lamellar forming polymer used here, we find $L_0= 42 \pm 0.7$~nm in agreement with other measurements~\cite{ji2008, Lee2005}.

\textbf{Mechanics}:  Once prepared, the composite sample is compressed with a custom built strain stage. When under compression, the film buckles out of plane forming a sinusoidal pattern in a process known as wrinkling.\cite{Bowden1998,Cerda2003,Groenewold2001,Huang2007,Stafford2004, Chung2011}.  The sinusoidal pattern itself is only stable in the small strain limit.  Applying a large strain\cite{Pocivavsek2008, Davidovitch2011, Holmes2010, Ebata2012} results in heterogeneous stress fields\cite{Witten2007, Tallinen2009}, which  lead to non-linear responses such as localized bending, shear deformation zones, crazes and delaminations\cite{Si2005, Lee2012, Mei2011, Audoly2010, Nolte2011}.  In this work we focus only on delamination that create free-standing film bends, however, other sharp features will also cause local plasticity.  Once delaminated, the film surface is imaged in three dimensions using LSCM, allowing an unambiguous determination of local curvatures.  When coupled with Euler-buckling theory, the curvature yields a value for the local bending induced strain. A typical experiment is shown in Fig.~\ref{before}. The mechanical compression is removed, and once again the film is imaged with LSCM or AFM (see Fig.~\ref{before}c and \ref{before}d).

\section{\textbf{Results and discussion}}

In a previous publication~\cite{Gurmessa2013}, Gurmessa and Croll reported the measurement of the onset strain for plasticity at the surface of a thin polystyrene film.  Several observations were made:  1.) the onset of plasticity occurs at low strain; scaling estimates put it at the order of $\sim 10^{-3}$, 2.) the critical strain for plastic failure remains constant for thicker films but rises as the film confinement increases.  3.) the confinement induced increases of the yield strain are independent of molecular weight and begin as the film thickness drops below about $\sim 100$~nm.  The observation of a low strain motivates the need to advance the modeling such that the strain estimate can be made quantitative, which need only involve a detailed exploration of model assumptions and the precise shape of the delaminations at their peaks.  Determining the generality of the second two observations is the main focus of the present work, and is accomplished through the use of additional polymer chemistry and architecture.  A second polymer must be chosen to have similar mechanical properties to polystyrene ensuring that no large-scale changes in experimental techniques are necessary.  For example, if a second polymer has a much smaller modulus than polystyrene an extreme strain on the substrate may be necessary to cause delamination.\cite{Ebata2012}  P2VP is an ideal molecule for such a comparison as its statistical segment size is almost identical to polystyrene, its bulk $T_{g}$ is similar and its bulk mechanical properties are also close to those of polystyrene.\cite{Argon1968, Kramer1974, Takahashi1996} 

\begin{figure}
\centering
\includegraphics[height=0.5\textwidth]{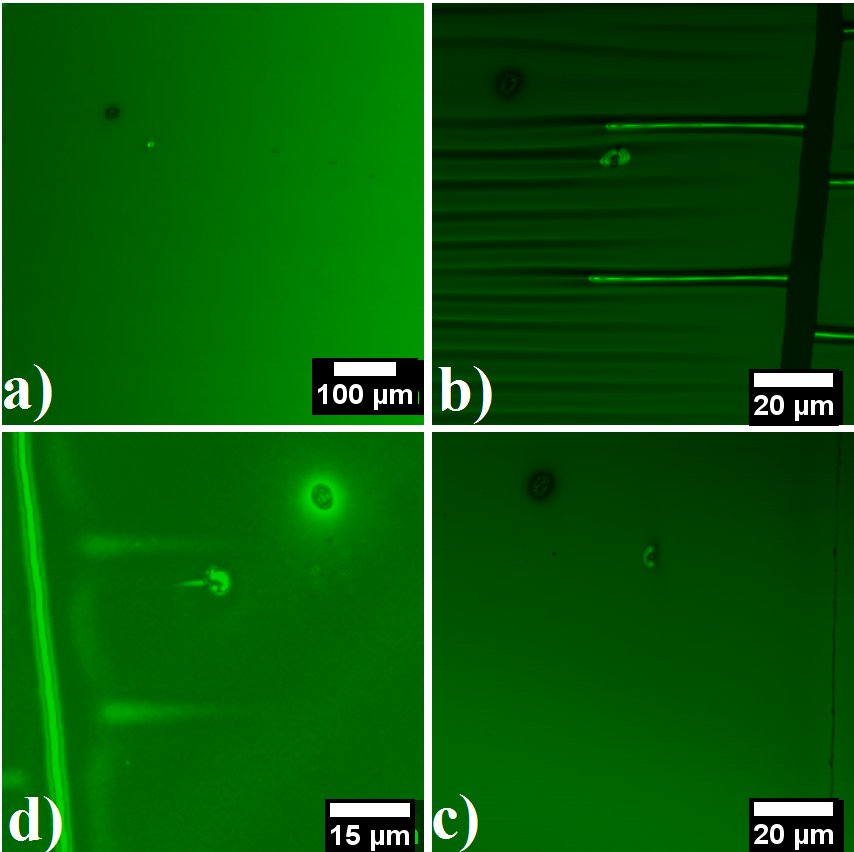}
\caption{ Typical surface morphologies of as cast PS-b-P2VP films at every stages of the mechanical loading. A confocal microscope image of (a) prestrain state  (b) the same film under compression, showing several wrinkles and delaminations (c) the relaxed film corresponding to the location of the previous delaminations and (d) the same film after it has been transferred to a silicon substrate.}\label{before}
\end{figure}

A typical experiment is illustrated in Fig.~\ref{before}.  In this case we show an as-cast film of block copolymer (which has more relevance to the discussion below).  The smooth film morphologically resembles what is observed with the P2VP homopolymer.  Specifically, the figure shows LSCM images of an as-cast  PS-b-P2VP film of thickness $h=71$~nm under:  zero-strain (\ref{before}a),  compression (\ref{before}b), removal of compression (\ref{before}c)  and after transferred to a highly reflective silicon substrate (\ref{before}d).  A smooth and flat surface is observed before strain is applied, but the film eventually evolves to wrinkles and delaminations (brightest wrinkle peaks) as compressive stress is increased.~\cite{Ebata2012}  Confocal  imaging of the sample while it is under compression, allows delamination width to be accurately measured.  More importantly, the heights of each pixel can be determined through LSCM's optical sectioning and the amplitude of the delamination can be determined.  

\begin{figure}
	\centering
	\begin{subfigure}[b]{0.49\textwidth}
		\includegraphics[height=.75\textwidth]{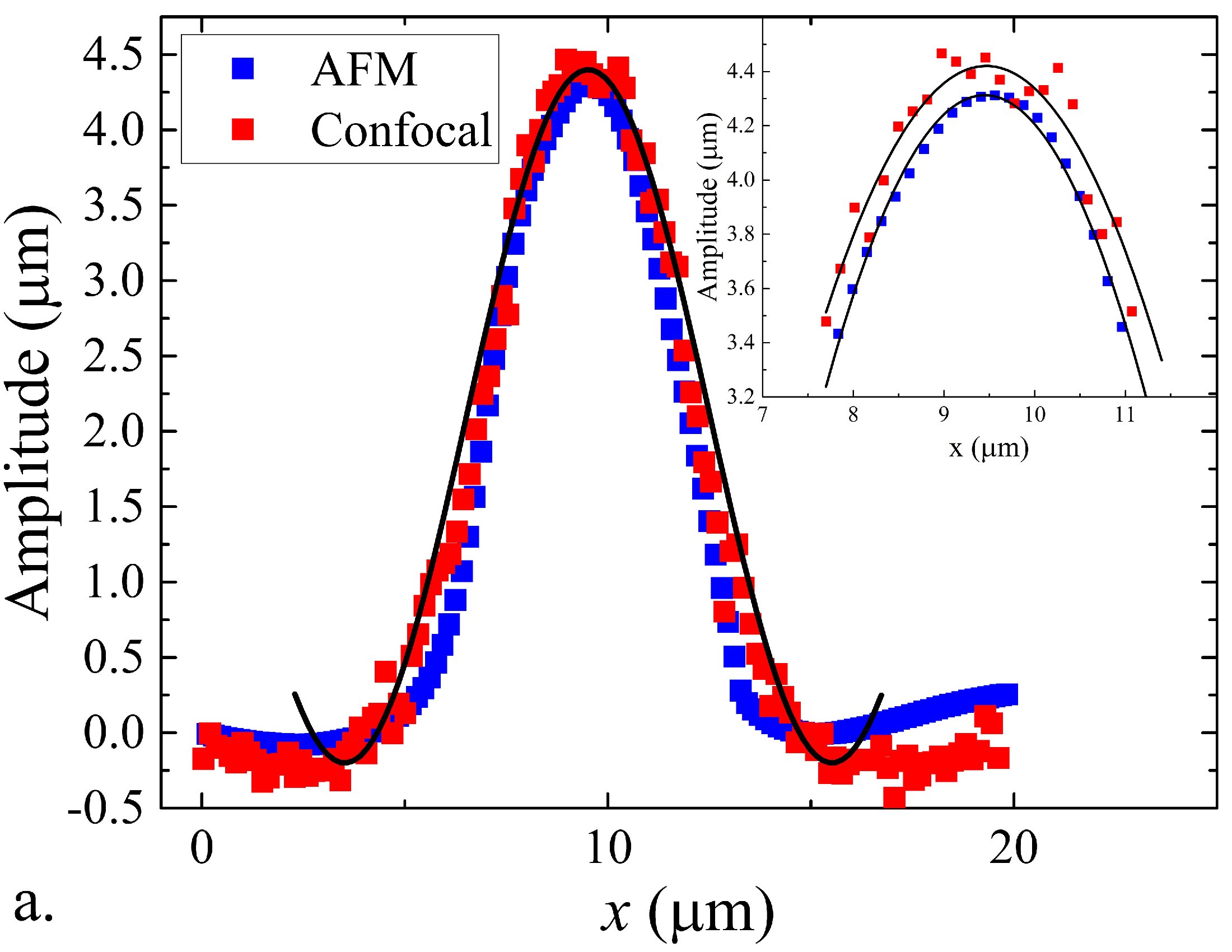}
	\end{subfigure}
	\begin{subfigure}[b]{0.49\textwidth}
		\includegraphics[height=.75\textwidth]{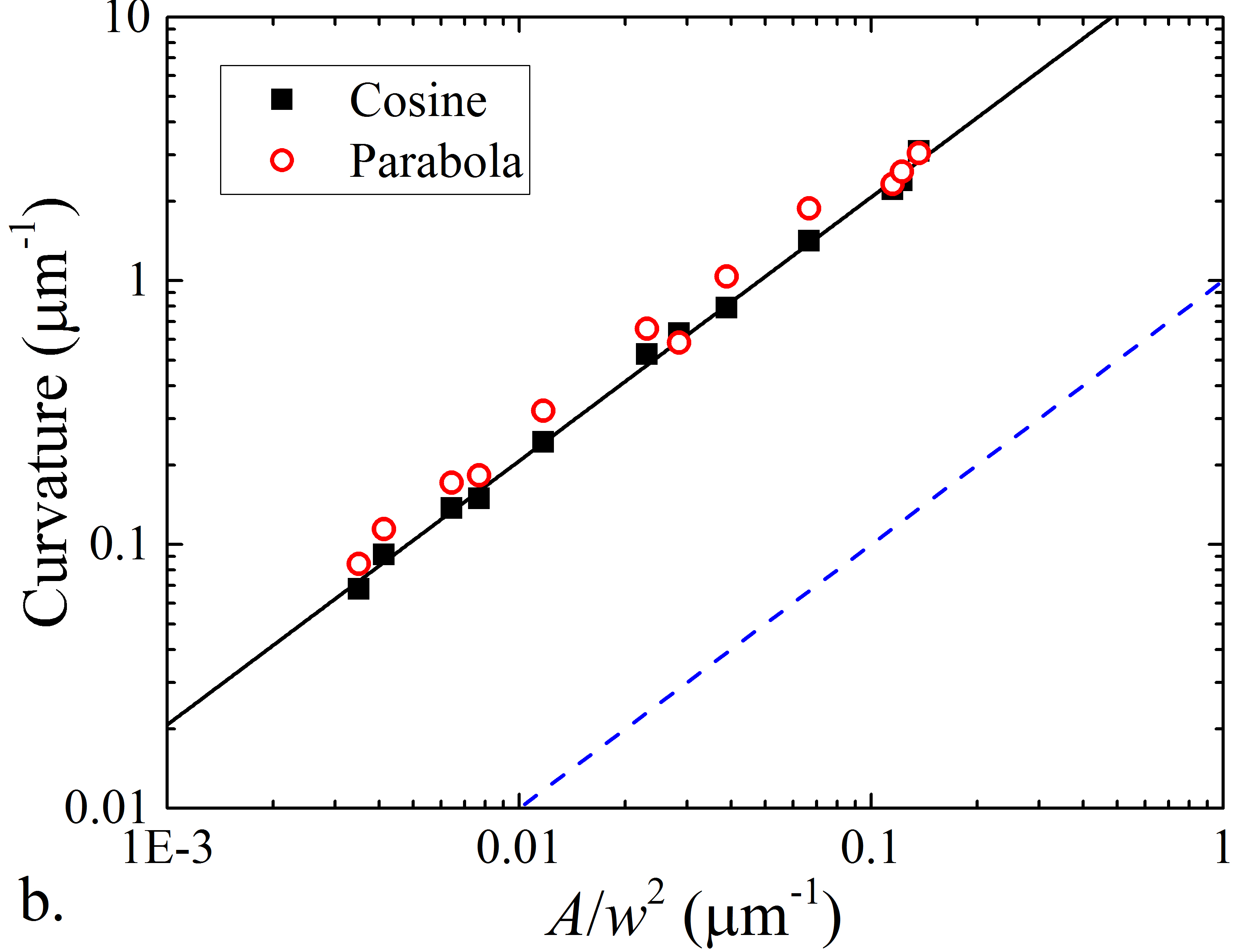}
	\end{subfigure}
\caption{(a) A typical delamination of a PS film measured by LSCM (red squares) and AFM (blue squares).  The film is $150$~nm thick.  The full delamination is fit with a cosine (solid black) while the peaks are fit with parabola (inset).  (b) Scaling plot of the peak curvature calculated from the cosine fit (black square) and the parabola fit (open circles).  Data are from PS, P2VP and as-cast PS-b-P2VP films of various thicknesses (ranging from $40$~nm to $315$~nm).  Solid line is a linear fit with a slope of $20.7$.  The dashed blue line shows a slope of $1$ for reference.   }\label{delam}
\end{figure}

Figure~\ref{delam} shows a typical delamination cross-section recorded with both AFM and Confocal microscopy.  The two measurements coincide within the error of locating a precise location along a delamination with two different instruments (essentially an error in length from a sample edge or other reference point to the location of interest).  Curvature of such a feature will scale as $\kappa\sim A/w^2$, where $A$ is amplitude, and $w$ is width as shown in Fig.~\ref{scheme1}.  Strictly speaking, this scaling  is only quantitatively correct if the delamination forms a perfect ellipse (which is clearly not the case).  In order to calculate a quantitatively accurate curvature, the precise shape of the delamination must be taken into account.  Recent theoretical predictions based on the F{\"o}ppl-von k{\'a}rm{\'a}n equation suggest the delamination takes the following form $g(x) = (A/2)(1+\cos (2 \pi x/ w))$, where $g(x)$ is the plate surface, $A$ is an amplitude and $w$ is the width of the delamination.~\cite{Vella2009}  The curvature is given by the second derivative of $g(x)$ evaluated at the peak of the delamination, $x=0$.  In this case $\kappa = 2\pi^{2}A/w^{2}$ and the scaling prefactor is numerically $19.74$.  A typical fit of $g(x)$ is shown in Fig.~\ref{delam}.  The fit is reasonable, but imperfect, due to imprecision in matching the minima.  In essence, the fit of such a function is dominated by the lateral location of the maxima and minima, not their respective vertical values.

A conceptually more accurate alternative is to fit the peak with a parabola and calculate the curvature from the parabolic fit ($h(x)=ax^2+bx+c$ and $\kappa \approx 2a$).  As shown in the inset of Fig.~\ref{delam}a the local fit is quite good.  Figure~\ref{delam}b shows a plot of the curvature of many different experimental delaminations determined by fits with $g(x)$ and $h(x)$ as a function of the scaling $A/w^2$.  Data points are take from PS, P2VP and as cast diblock films, and thicknesses range from $40$~nm to $315$~nm.  We note no deviation from either curve for any of the data - this should be expected for a purely geometric feature such as the curvature.   Finally, both curves are linear (as one would expect) and have similar slopes.  The hypothesis that the curvature is better represented by the parabolic fits is not born out in practice.  The measured slope is $20.7\pm .5$.  All data discussed below is adjusted to a quantitative scale using this measured prefactor.

\begin{figure}
\centering
\includegraphics[width=\textwidth]{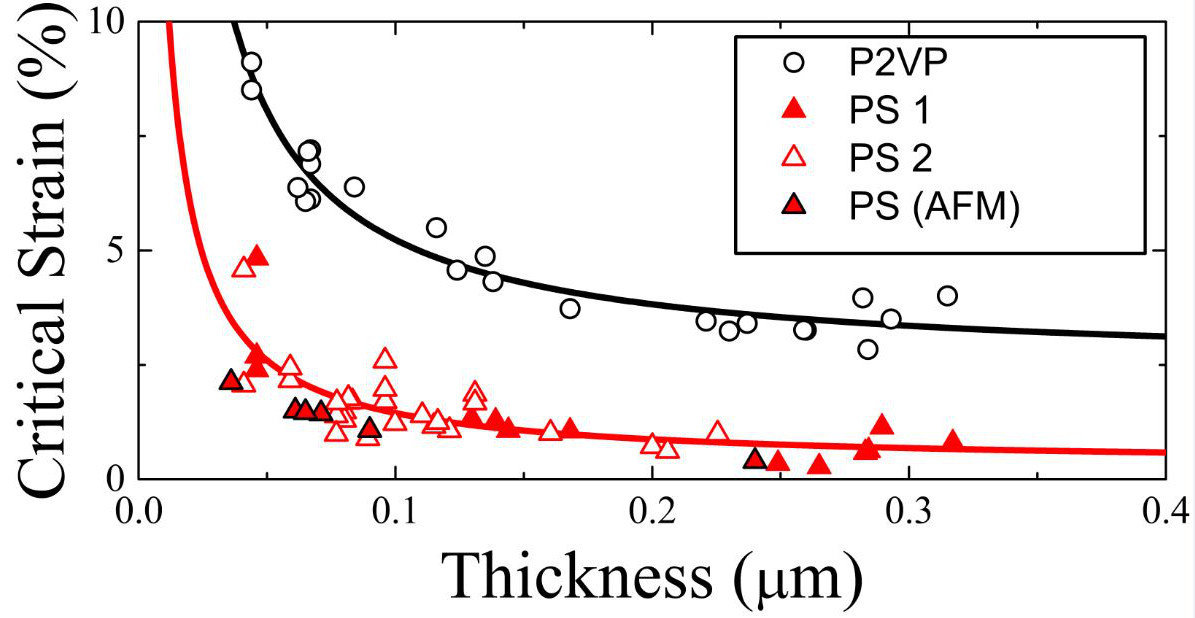}
\caption{ Comparison of the evolution of yield stain of PS and P2VP homopolymer as a function of confinement. Both polymers show increases in the critical strain for plasticity as thickness falls below $100$~nm.  Solid curves are fits to the layer model discussed in the text.}\label{P2VP_PS}
\end{figure}

There is a strong correlation between high delamination curvature and the the amount of damage observed once film is transferred to a flat silicon substrate.  For example, much of the film which only displayed low curvature features (such as the wrinkles) while deformed show no signs of damage at the end of the experiment, whereas regions of high curvature (tall delaminations) show clear signs of damage.  The critical point for plastic deformation, the point of lowest curvature which still results in damage, can easily be extracted from the data and the curvature at this point can be calculated.  The surface strain experienced at this location is given by $\epsilon = h\kappa/2$, where $h$ is the film thickness.  Fig.~\ref{P2VP_PS} shows the critical strain for plasticity as a function of film thickness for both PS (the same data as in reference \cite{Gurmessa2013} adjusted with the scaling prefactor discussed above) and homopolymer P2VP.

The behavior of P2VP largely mirrors that of PS films.  Both polymers have low critical strains in the traditional `thick' film region, but show strains that dramatically increases as confinement increases.  Notably, P2VP has critical strains $\sim$ 2\% greater than PS over the entire range of measurement as illustrated in~Fig \ref{P2VP_PS}.  The result shows that the critical strain measured here is indeed a material dependent property and not a geometric artifact.  Also, the slightly higher values found for P2VP mirror the differences in their traditionally measured bulk values.\cite{Takahashi1996}  Finally, the increased threshold in PS and P2VP is consistent with the changes observed in the glass transitions of free-standing films of the two materials, suggesting that at least a qualitative connection may exist between $T_{g}$ and the plasticity we observe.\cite{Keewook2011}

To aid in interpreting the data, fits are made to a simple layer model, similar to that proposed by Keddie~\cite{Keddie1994} to explain the thickness dependence of glass transition temperature of polymer thin  films.  The model assumes an enhanced molecular mobility (a liquid-like layer) near the free-surface, compared to the bulk polymer.  Several authors~\cite{Keddie1994, Fakhraai2008, Ellison2003, Pye2011, OConnell2005, Bohme2002} have shown that the glass transition temperature of thin polystyrene films is reduced significantly compared to the bulk value as confinement increases.  The depression in $T_g$ is attributed to the enhanced mobility of chain segments residing at the free surface which is caused by the reduction of barriers to segmental cooperative motion\cite{Roth2007a}.  Here the layering will give rise to different yield strains at the surface and in the bulk.  Specifically, the yield strain as a function of thickness for a two layered system is, 
\begin{equation}\label{Layer}
\epsilon_p (h) = (\ell / h) ( \epsilon _{p} - \epsilon _{p} ^{0}) + \epsilon _{p} ^{0},
\end{equation}
where $\ell$ is the size of the soft layer and $\epsilon _{p}$ and $\epsilon _{p} ^{0}$ refer to the surface and bulk strain, respectively.  A larger strain is needed in the liquid-like layer before stress can be stored; the stress can more easily relax away in the liquid layer than in the bulk layer.  Fitting to the data in Fig.~\ref{P2VP_PS} give $\epsilon _{p} = 12 \pm 1$\%, $\epsilon _{p} ^{0} = 0.3 \pm .2$\%,  and $\epsilon _{p} =  31 \pm 1$\%, $\epsilon _{p} ^{0} = 2.4 \pm .1$\%, respectively, for PS and P2VP, assuming a typical lengthscale $\ell = 10$~nm.\cite{Keewook2011}  

The plateau values are smaller than yield strains measured in bulk, but this is not unexpected.  The technique used here focuses on local, microscopic signs of damage.  This kind of sensitivity is not possible with a bulk sample, because only force and displacement are measured in a traditional bulk experiment.  A sample must move enough material plastically to be measured as hysterisis in a force-displacement curve.  Furthermore, it would be extremely difficult to image the beginnings of plastic rearrangement after a bulk experiment because the entire bulk sample would need to be examined.

The block copolymer poses new challenges to interpretation because of its ability to microphase separate.  In the disordered state the two blocks are well mixed and (upon vitrification) the solid is easily modeled as a linear, isotropic, continuum material.  As the material begins to phase separate (or fluctuations become large) it is no longer clear that the material can be thought of in a continuum sense.  However, with the small size of the domains and the long range of mechanical perturbation, continuum ideas are often applied successfully to microphase separated solids.\cite{ Schwier1985, Honeker1996, Weidisch1999a, Weidisch1999b, Weidisch2000, Cohen2000, Lach2004, Ye2013, Ryu2002, Ruokolainen2002, Lee2005, Kim2006}  We follow similar assumptions, and make no changes to our modeling of the basic mechanical deformation (e.g. calculation of a surface strain) after microphase separation has taken place.

In order to follow the microstructures influence on the onset of plasticity in PS-b-P2VP films, the self-assembly process of the diblock copolymer microstructure must be carefully controlled.  Self-assembly is typically carried out through thermal annealing~\cite{Bates1996, Ruokolainen2002, Mai2012} or solvent annealing~\cite{Grozea2011}.  Solvent annealing may leave residual stresses when the sample is quenched, but these stresses can be minimized with additional thermal annealing.  In a typical thermal experiment,  the copolymer is heated past its glass transition temperature, ($T_g \sim100$~$^{\circ}$C for PS and P2VP),  for a predetermined time followed by rapid quenching of the sample to room temperature (below $T_g$).  When the sample is quenched below its glass transition temperature, the polymer structure is kinetically frozen due to the extremely low mobility of the chains.  Given the composition of the block copolymer considered (PS block of $M_n = 40$~kg/mol and a P2VP block of $M_n = 40.5$~kg/mol), lamella parallel to the mica substrate are expected at equilibrium.~\cite{ji2008}  In this system, the PS block has a lower surface energy (compared to P2VP) and will reside at the free surface whereas P2VP favors the substrate, yielding lamella parallel to the substrate.  

\begin{figure}
\centering
\includegraphics[width=0.8\textwidth]{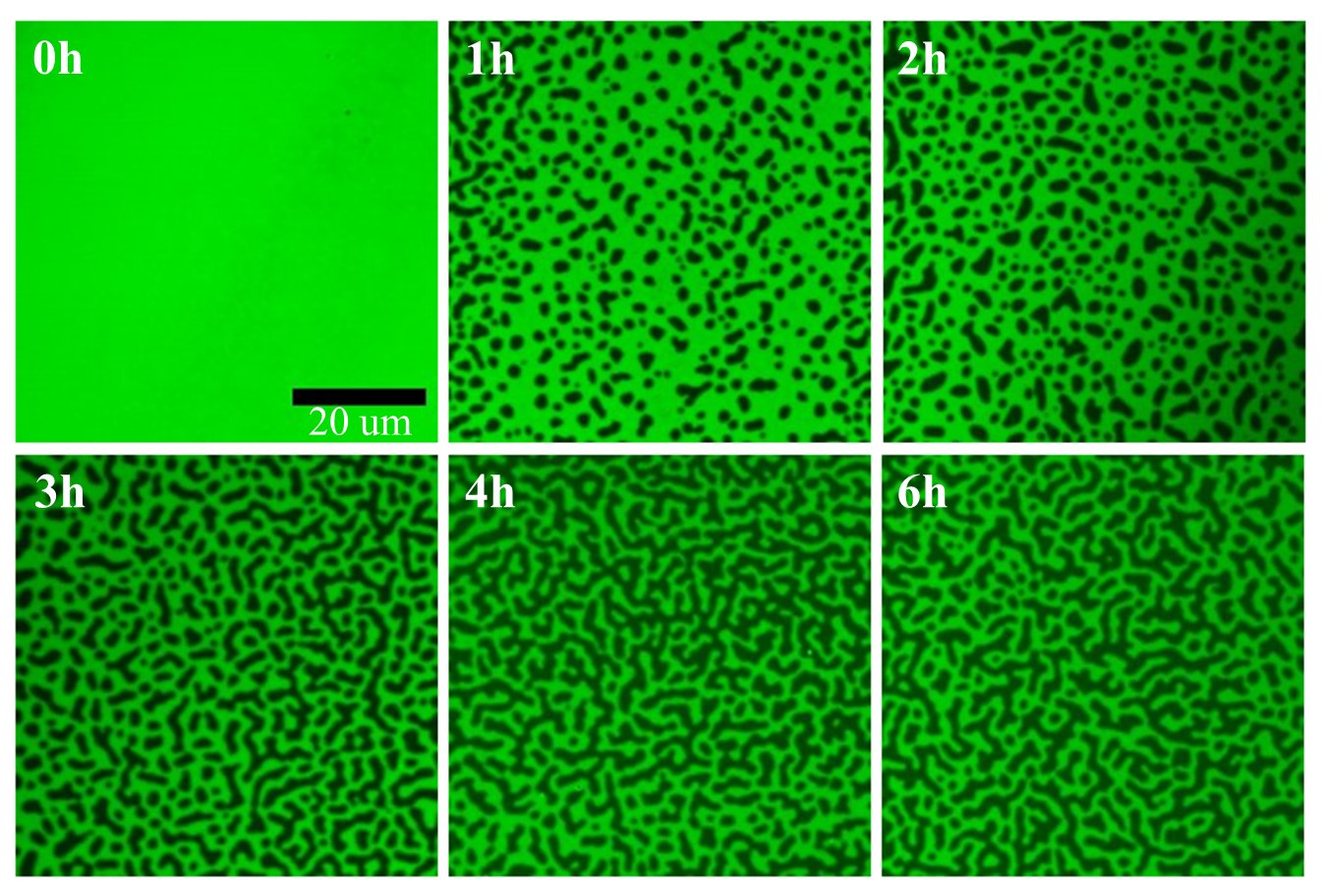}
\caption{Surface morphologies of samples annealed at $175$~$^\circ$C for the annealing times indicated.  All of the samples are decorated by holes of lamella spacing $\sim42$~$\mu$m.  Scale bar indicates $20$~$\mu$m and the sample at T=0 has a thickness of $125 \pm 5$~nm.}\label{TEvoln}
\end{figure}

A series of samples  of similar thicknesses were annealed to temperatures ranging from $165$~$^{\circ}$C - $195$~$^{\circ}$C under variable environments (air or a glove box filled with dry nitrogen).  When annealed to a temperature above the glass transition for an appropriate time, the lamellar structure will force the surface of the block copolymer thin film to be decorated by terraces (holes, islands, bi-continuous patterns).  Fig.~\ref{TEvoln} illustrates the surface morphological evolution observed as the samples of similar thicknesses that are annealed at a temperature of $175$~$^{\circ}$C for different annealing times.  As time progresses, the domain size coarsens as expected.  Fig.~\ref{TEvoln} makes it apparent that self-assembly of the polymer chains into periodic microdomains occurs at relatively short annealing times.  It is important to note that the appearance of a terraced morphology does not necessarily mean that microphase separation has completed, or that material properties have stopped changing.~\cite{Mayes1994, Croll2009}

\begin{figure}
\centering
\includegraphics[width=0.5\textwidth]{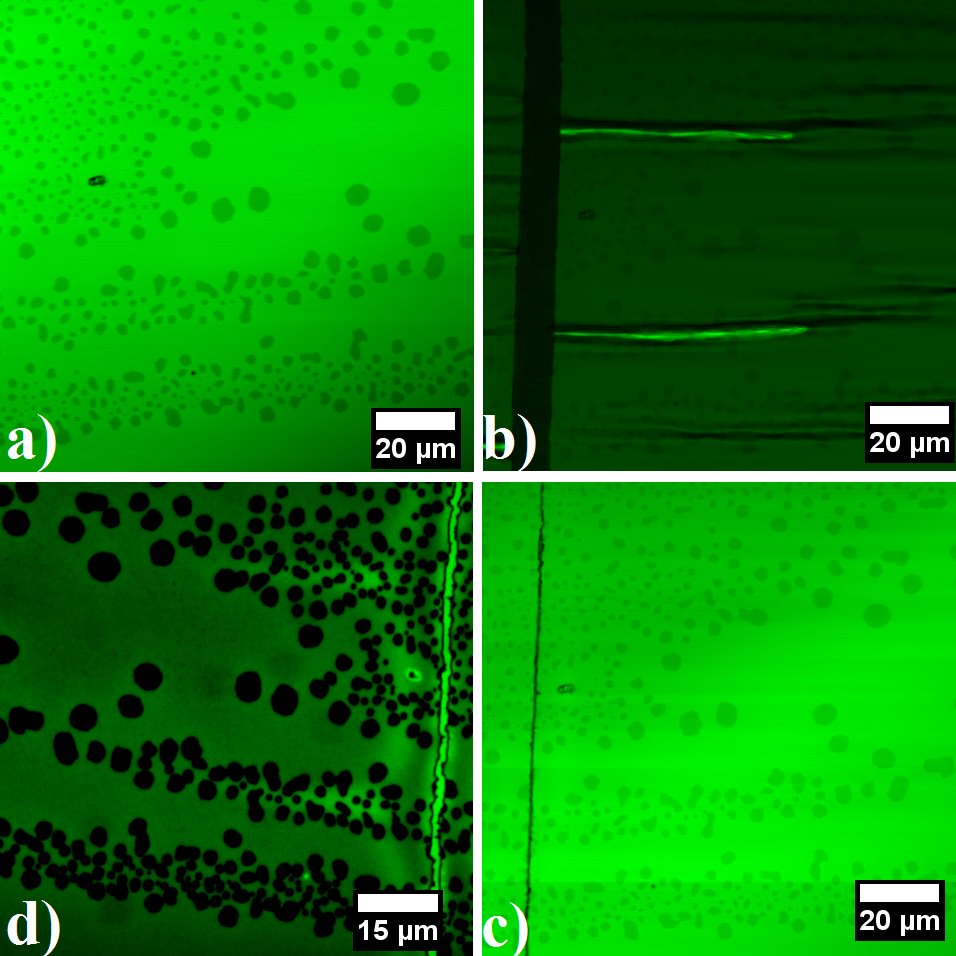}
\caption{Surface morphologies of ordered PS-b-P2VP films at various stages of the experiment. A confocal microscope image of a.) prestrain state.  b.) the same film under compression with several wrinkles and delaminations.  c.) the same location of the film after removal of compression.   d.)  the same film after it has been transferred to silicon. }\label{after}
\end{figure}

\begin{figure}
\centering
\includegraphics[width=0.5\textwidth]{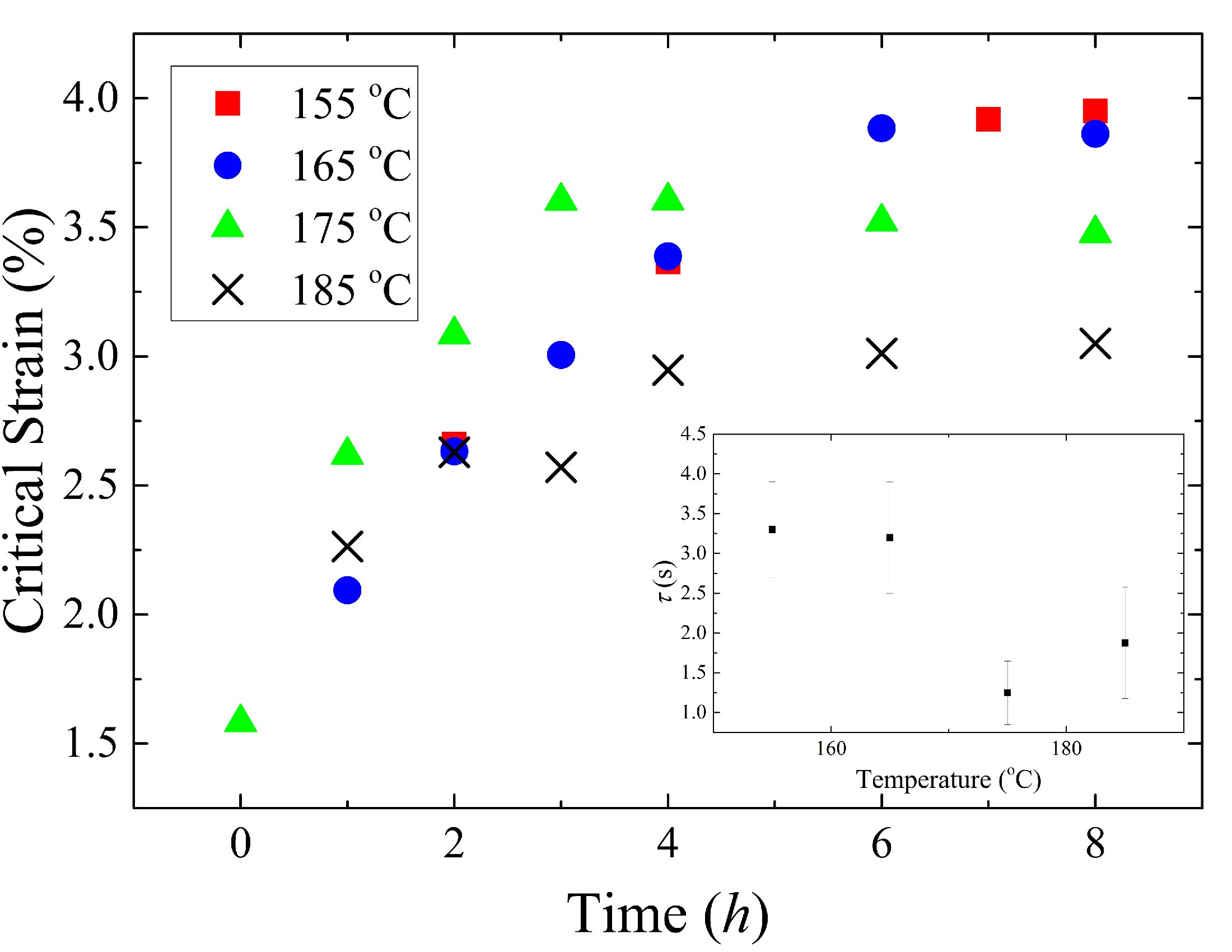}
\caption{Evolution of critical strain in block copolymer films as the annealing time increases. The figure shows that the yield strain, regardless of annealing temperature, equilibrates to a finite value after a relatively short annealing time. The inset shows a plot of the time constant, $\tau$,  as a function of annealing temperature.}\label{evolution}
\end{figure}

Tests of the critical strain were then conducted with microphase separated PS-b-P2VP samples using the same procedures outlined above. The pre and post buckling state of the sample are illustrated in Fig.~\ref{after}.  The location of the delaminations is easy to follow through each stage of the experiment due to the unique pattern formed by the terraces (note that the sample in Fig.~\ref{after} is of non-uniform thickness and is only used to highlight the comparison).  Similar to the as cast films, a flat surface is observed when no strain is applied, and the film eventually evolves to wrinkles and delaminations (brightest wrinkle peaks) as global compressive stress is increased.  The sample once again flattens as the compression is removed.  The experiment was repeated for as-cast films and samples of identical thickness treated to annealing at various temperatures and times.  The results are summarized in Figure~\ref{evolution}.
 
For all temperatures qualitatively similar behavior is observed.  The critical strain for plasticity increases as annealing takes place until it reaches a plateau value which does not change significantly upon further annealing.  The sample annealed at $175$~$^\circ$C reaches the plateau after four hours of annealing, notably longer than it takes for the surface to break up into well defined islands (compare with Fig.~\ref{TEvoln}).  Unfortunately the sample to sample variation obscures the fine details of the annealing process.  For example, under the assumption of a single time constant exponential process (e.g. $\epsilon (t) \sim A-Be^{-t/\tau}$) no clear trend is visible (see the inset of Fig.~\ref{evolution}).  The variation is likely due to small thickness differences between sets of samples (we measure of order 10 nm in thickness variation by AFM) coupled with the sensitivity of the island formation process to film thickness.\cite{collin1992}  Regardless, the overall trend displayed in Figure~\ref{evolution} does allow two distinctly different regimes to be identified: as-cast and well annealed.

\begin{figure}
	\centering
	\begin{subfigure}[b]{0.5\textwidth}
		\includegraphics[height=.2\textheight]{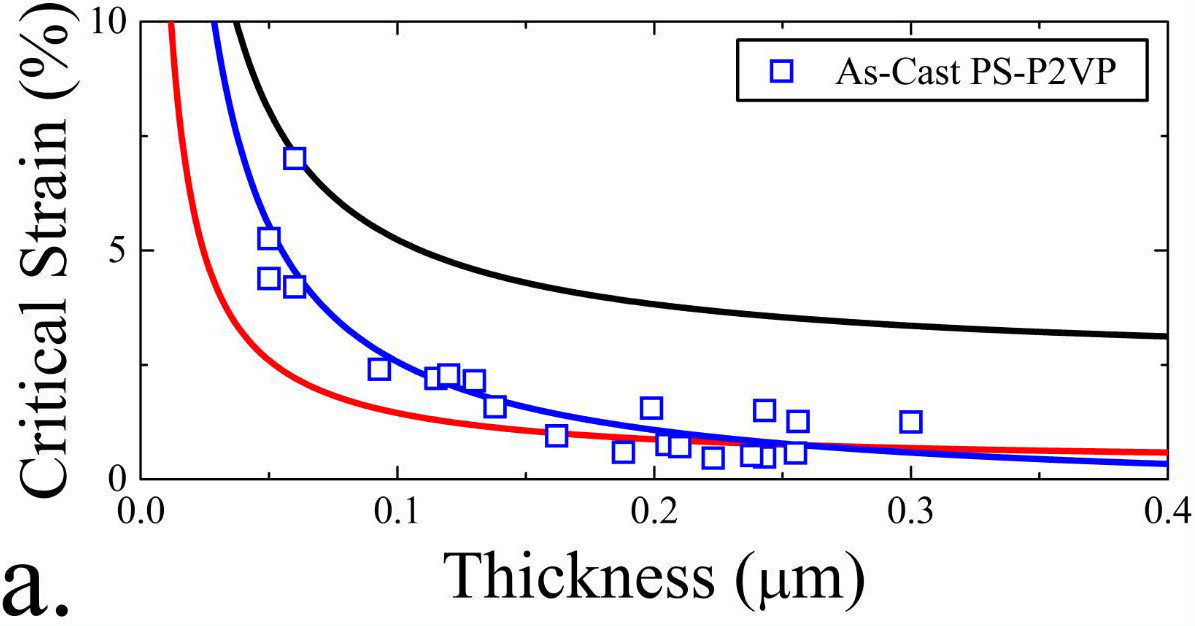}
\hspace{5em}
	\end{subfigure}
	\begin{subfigure}[b]{0.5\textwidth}
		\includegraphics[height=.2\textheight]{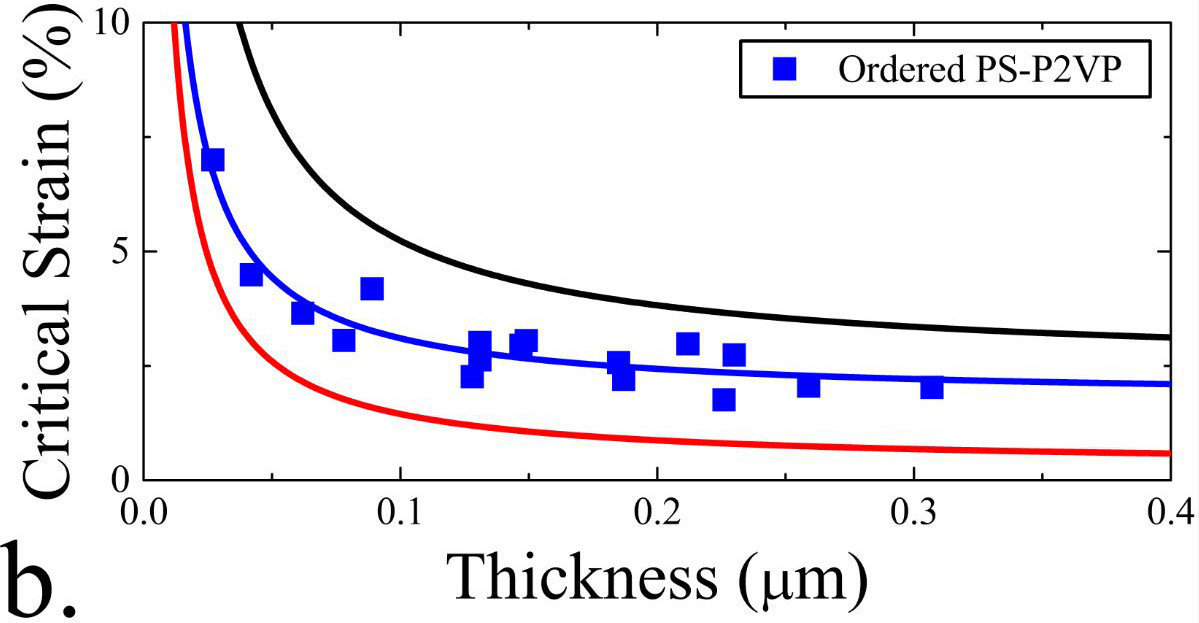}
	\end{subfigure}
\caption{Critical strain as a function of film thickness of as cast and PS films fitted to the layer model described in the text.   \label{diblock}}
\end{figure}

As-cast PS-b-P2VP films of varying thickness were constructed and the critical strain for plasticity was measured.  Figure~\ref{diblock}a shows the result, along with the layer model fits to the PS and P2VP homopolymer.  The as-cast films have a 'bulk' value similar to PS but as the films become thinner they fall off the PS trend-line and show a greater than PS change in critical strain.  Fitting with the layer model, maintaining the same $10$~nm surface layer, results in unphysical strain values ($\epsilon _{p} = 29 \pm 3$\%, $\epsilon _{p} ^{0} = -0.4 \pm .3$\%).  The reason for the complex behavior is rooted in the non-equilibrium nature of the as-cast state.  As a film is spin cast, polymer interacts with the two surfaces (here air and mica) until solvent evaporates and the film vitrifies.  This means that there is often some degree of order present after spin casting a film simply because of the surfaces.\cite{Mayes1994}  Similar to other surface effects, the ordering will disproportionately affect very thin films.  In this case, the thinnest films are largely ordered at the end of the spin coating process.  A fact easily verified by repeating the measurement with annealed samples of similar thickness.

Figure~\ref{diblock}b. shows the results after long annealing times ($> 4$~h).  The trend now is found roughly halfway between the PS homopolymer data (red curve) and the P2VP data (black curve).  The thinest films show critical strains that are identical to the values of the as-cast films, verifying our earlier hypothesis.  A fit to Eqn.~\ref{Layer} yields $\epsilon _{p} = 15 \pm 1$\%, $\epsilon _{p} ^{0} = 1.8 \pm .2$\%.  We tentatively interpret the result as an indication that a simple `mixing-rule' can be applied to the annealed samples; they display properties proportionately to the volume fraction of each material used to make them up.  While this may be true for the relatively isotropic lamellar forming system used above, measurements conducted with a cylinder forming PS-P2VP molecule (of volume fraction $f=.37$ and similar total molecular weight and film thickness) show very little change from the as cast to ordered state (the ratio of critical strains is measured to be $\sim0.95$).  Taken together, the conclusion can only be that long range connectivity must play a major roll in the mechanics of these thin films because the lamellar film has long range order, whereas the cylinder forming and as-cast material does not.

In summary, the critical strain for plasticity was measured in PS, P2VP and block copolymers of PS and P2VP.  Contrary to earlier work, the analysis presented here moves beyond scaling estimates and is put into a quantitative framework.  For all polymer studied the critical strain was found to reach a constant but material dependent value in thick films.  Specifically, thick P2VP homopolymer becomes irreversibly deformed at a strain of $\epsilon = 2.4$~\%, PS homopolymer at a strain of $\epsilon = 0.3$~\%.  As-cast symmetric diblock copolymer has a polystyrene like value, whereas well ordered block copolymer has a critical strain of $\epsilon =1.8$~\%.  PS rich cylinder forming block copolymer was found to have a polystyrene-like critical strain, indicating that long range order plays a roll in the deformation process.  More interestingly, all polymer shows an increase in critical strain as thickness falls below $\sim100$~nm, consistent with an assumed relation between the glass transition and plasticity in thin polymer films.

\begin{acknowledgement}

B.G. thanks NSF EPSCOR for a doctoral dissertation assistantship and A.B.C. thanks NSF EPSCoR (EPS-0814442) for generous support and R. Huang for useful discussion.

\end{acknowledgement}

\bibliography{DTFB}

\end{document}